\documentclass{amsart}
\usepackage{amssymb}
\usepackage{amsfonts}
\usepackage{graphicx}
\usepackage{epstopdf}
\usepackage{hyperref}

\newcommand{\beq}{\begin{equation}}
\newcommand{\eeq}{\end{equation}}

\begin{document}

\begin{center}
{\Large \bf Partition function of Chern-Simons theory as renormalized q-dimension \\
\vspace*{1 cm}

{\large  R.L.Mkrtchyan\footnote{mrl55@list.ru}
}
\vspace*{0.2 cm}

{\small\it Yerevan Physics Institute, 2 Alikhanian Br. Str., 0036 Yerevan, Armenia}

}

\end{center}\vspace{2cm}

{\small  {\bf Abstract.}We calculate  $q$-dimension of $k$-th Cartan power of fundamental representation $\Lambda_0$, corresponding to affine root of affine simply laced Kac-Moody algebras, and show that in the limit $q\rightarrow 1 $, and with natural renormalization, it is equal to universal partition function of Chern-Simons theory on three-dimensional sphere. }



\section{Introduction}

In this note we present an observation, which provides another type of well-known connection of Chern-Simons theory and Kac-Moody affine algebras.  We postpone  discussion of this observation to Conclusion, and below in Section 1  first consider partition function of Chern-Simons theory on three-dimensional sphere \cite{Wit1}. We present it in the special (universal) form \cite{M13,KM}, then in Section 2  calculate the $q$-dimension \cite{Kac} of highest weight representation with highest weight $k \Lambda_0, k \in  \mathbb{Z}_+$ of affine Kac-Moody algebras, and in Section 3 show that these expressions have  coinciding integrands and coincide exactly after natural renormalization of latter in $q\rightarrow 1$ limit. In Conclusion we also discuss possible applications.

\section{Universal representation of partition function of Chern-Simons theory on 3d sphere}

Partition function of Chern-Simons theory on 3d sphere was first calculated in \cite{Wit1} to be $S_{00}$ element of the matrix of modular transformations. In \cite{M13} it was represented in universal form:

 \begin{eqnarray}\label{totalfree0}
\mathcal{F} =\frac{d}{2}\ln(y/t) + \int^{\infty}_0 \frac{dx}{x} \frac{f(x/y)-f(x/t)}{e^{x}-1} 
 \end{eqnarray}

where $d$ is dimension of gauge group:

\begin{equation} \label{dim}
d =\frac{(\alpha-2t)(\beta-2t)(\gamma-2t)}{\alpha\beta\gamma}\,,
\quad (t=\alpha+\beta+\gamma)\,.
\end{equation}

and $f(x)$ is character of adjoint representation, restricted on Weyl line: 

\begin{eqnarray}
 \chi_{ad}(x\rho)&=& f(x)\\
 f(x)&=& \frac{\sinh(x\frac{\alpha-2t}{4})}{\sinh(x\frac{\alpha}{4})}\frac{\sinh(x\frac{\beta-2t}{4})}{\sinh(x\frac{\beta}{4})}\frac{\sinh(x\frac{\gamma-2t}{4})}{\sinh(x\frac{\gamma}{4})}
 \end{eqnarray}
where Weyl line  $x\rho$ in the root space is one-dimensional subspace along Weyl vector $\rho$, which is the sum of fundamental weights of a given simple Lie algebra.  Vogel's (or universal) parameters $\alpha, \beta, \gamma$ correspond to different simple Lie algebras according to Vogel's table \cite{V}. Finally, $y=\kappa + t$ is shifted in a standard way Chern-Simons coupling constant $\kappa$, appearing in front of Chern-Simons action. Note also that, as usual in universal formulae, they are written with arbitrary normalization of invariant bilinear form in corresponding simple Lie algebra. In the so-called minimal normalization, when the square of long root is equal to 2, we recover usual formulae, e.g. for Chern-Simons theory $\kappa$ becomes an integer coupling $k$ (level). As shown in \cite{KM}, expression (\ref{totalfree0}) makes sense for values of universal parameters out of Vogel's table, also. Particularly, it contains non-perturbative corrections to the Gopakumar-Vafa partition function of dual topological string.

Introducing $q= e^x$, we have

 \begin{eqnarray}
f(x)= \frac{\left( q^{\frac{\alpha-2t}{4}}- q^{-\frac{\alpha-2t}{4}} \right)\left( q^{\frac{\beta-2t}{4}}- q^{-\frac{\beta-2t}{4}} \right)\left( q^{\frac{\gamma-2t}{4}}- q^{-\frac{\gamma-2t}{4}} \right)}{\left(q^{\frac{\alpha}{4}} -q^{-\frac{\alpha}{4}} \right)\left(q^{\frac{\beta}{4}} -q^{-\frac{\beta}{4}} \right)\left(q^{\frac{\gamma}{4}} -q^{-\frac{\gamma}{4}} \right)}, x=\ln q
\end{eqnarray}

We transform expression for free energy by introduction of function $F(x)$:

\begin{eqnarray}
F(x)=f(x)-d
\end{eqnarray}
which is $ O(x^2)$ at $x\rightarrow 0$, so integrals below converge. Cancellation in the last lines of (\ref{totalfree1}) and correspondingly representation (\ref{totalfree2}) are first observed in \cite{KM}.

 \begin{eqnarray}\label{totalfree1}
\mathcal{F} =\frac{d}{2}\ln(y/t) + \int^{\infty}_0 \frac{dx}{x} \frac{F(x/y)}{e^{x}-1} - \int^{\infty}_0 \frac{dx}{x} \frac{F(x/t)}{e^{x}-1} = \\ \nonumber
\frac{d}{2}\ln(y/t) + \int^{\infty}_0 \frac{dx}{x} \frac{F(x)}{e^{xy}-1} - \int^{\infty}_0 \frac{dx}{x}\frac{F(x)}{e^{xt}-1} = \\  \nonumber
\frac{d}{2}\ln(y/t) + \int^{\infty}_0 \frac{dx}{x} F(x) \left( \frac{1}{e^{xy}-1} -  \frac{1}{e^{xt}-1} \right) = \\ \nonumber
\frac{d}{2}\ln(y/t) + \frac{1}{2} \int^{\infty}_{-\infty} \frac{dx}{x} F(x) \left( \frac{1}{e^{xy}-1} -  \frac{1}{e^{xt}-1} \right) = \\ \nonumber
\frac{d}{2}\ln(y/t) + \frac{1}{2} \int_{R_+} \frac{dx}{x} f(x) \left( \frac{1}{e^{xy}-1} -  \frac{1}{e^{xt}-1} \right) -\\ \nonumber
\frac{1}{2} \int_{R_+} \frac{dx}{x} d \left( \frac{1}{e^{xy}-1} -  \frac{1}{e^{xt}-1} \right) = \\  \nonumber
 \frac{1}{2} \int_{R_+} \frac{dx}{x} f(x) \left( \frac{1}{e^{xy}-1} -  \frac{1}{e^{xt}-1} \right)
 \end{eqnarray}
where $R_+$ is slight deformation of $R$ (real line of variable $x$), bypassing singularity at zero from above (equally well we might use contour  $R_-$, since residue at $x=0$ is zero due to even integrand). So finally we have 

 \begin{eqnarray}\label{totalfree2}
\mathcal{F}  = \label{CS}
 \frac{1}{2} \int_{R_+} \frac{dx}{x} f(x) \left( \frac{1}{e^{xy}-1} -  \frac{1}{e^{xt}-1} \right)
 \end{eqnarray}

This expression will be compared with the purely representation-theory object, calculated below in Section 3.

\section{$q$-dimension of $k\Lambda_0$ representation of Kac-Moody algebras}

In this Section we calculate so-called $q$-dimension of special representations of affine Kac-Moody algebras, dual to untwisted ones. 

Basic objects in definition of affine Kac-Moody algebra $g(A)$ are two linearly independent sets: that of simple roots $\alpha_i \in \mathfrak{h}^*$ ($\mathfrak{h}^*$ is an $(l+2)$-dimensional root space) and coroots $\alpha_i^{\vee} \in \mathfrak{h}$ ($\mathfrak{h}$ is an $(l+2)$-dimensional Cartan subalgebra), $i=0,1,...,l$, and $(l+1)\times(l+1)$ generalized Cartan matrix $A$ of rank $l$ with matrix elements $a_{ij}$. One of the basic relations is

\begin{eqnarray}
\langle \alpha_i^{\vee}, \alpha_j \rangle =a_{ij}, \,i,j=0,1,...,l
\end{eqnarray}

For references to affine Kac-Moody algebras see e.g. \cite{Kac,Goddard:1986bp}, our notations exactly follow those of \cite{Kac}.

Consider highest-weight representation $V$ of affine Kac-Moody algebra $g(A)$ with highest weight $\Lambda \in \mathfrak{h}^*$  \cite{Kac}. Weight spaces decomposition of $V$ is  

\begin{eqnarray}
V= \underset {\lambda \leq \Lambda}{\oplus} V 
\end{eqnarray}

Defining {\it{degree}} of weight  $\lambda=\Lambda -\sum_i k_i \alpha_i, k_i \in \mathbb{Z}_+$ in $V$ as

\begin{eqnarray}
\text{deg}(\lambda)=\text{deg}(\Lambda -\sum_i k_i \alpha_i)=\sum_i k_i
\end{eqnarray}

one introduce {\it{principal gradation}} (\cite{Kac}, \S 10.10) of V:

\begin{eqnarray}
V & = & \underset {j \geq 0}{\oplus} V_{j} \\
V_j & = & \underset {\lambda: \text{deg}\lambda=j}{\oplus} V_{\lambda}
\end{eqnarray}
where sum is over non-negative integers $j$, and define $q$-{\it{dimension}} of $V$:

\begin{eqnarray}
\text{dim}_q (\Lambda)=\sum_{j \geq 0} (\text{dim}V_j) q^j
\end{eqnarray}

The product representation of $\text{dim}_q(\Lambda)$ can be derived (\cite{Kac}, 10.10.1) from general Kac-Weyl formula for characters and is given by

\begin{eqnarray} \label{qd}
\text{dim}_q(\Lambda)=\prod _{\alpha\in \Delta_+^{\vee}} \left( \frac{1-q^{\langle \Lambda+\rho,\alpha\rangle}}{1-q^{\langle \alpha,\rho\rangle }} \right)^{ \text {mult}(\alpha)}
\end{eqnarray}

where $\rho \in \mathfrak{h}^*$ is defined by relations

\begin{eqnarray}
\langle \rho, \alpha_i^{\vee} \rangle & =& 1, i=0, ..., l \\
\langle \rho,d \rangle &=& 0
\end{eqnarray}
and product is over $\alpha \in \Delta_+^{\vee} $ - positive roots of dual Kac-Moody algebra. Here $d$ is defined \cite{Kac} as an arbitrary element of $\mathfrak {h}$ satisfying $\langle \alpha_0,d\rangle=1, \langle \alpha_i,d\rangle=0, i=1,...l$.

We would like to consider representations with $\Lambda$  proportional to fundamental weight of the affine root: $\Lambda \rightarrow \lambda_k = k \Lambda_0, k \in Z_+$, where $\Lambda_0$ is defined by relations

\begin{eqnarray}
\langle \Lambda_0,\alpha_i^{\vee}\rangle &=& 0, i=0,1,...,l \\
\langle \Lambda_0,d\rangle &=& 0
\end{eqnarray}

$\rho$ can be represented as

\begin{eqnarray}
\rho= \bar {\rho} +h^{\vee}\Lambda_0
\end{eqnarray}
where $h^{\vee}$ is dual Coxeter number and $\bar {\rho}$ is orthogonal projection of $\rho$ on the Cartan subspace of roots of underlying simple Lie algebra (which is linear span of $\alpha_i, i=1,2,...,l)$, i.e. essentially the Weyl vector of underlying simple Lie algebra, Dynkin diagram of which is obtained by removing 0-th node of initial Dynkin diagram of affine Kac-Moody algebra. 

We consider from now on simply-laced  untwisted affine algebras, i.e.       $\hat{A} \hat{D}\hat{E}$ algebras. Roots of untwisted algebras can be described as follows. Real roots (hence of multiplicity one) for affine Kac-Moody algebra are given by $ \{ \alpha +n \delta| \alpha \in \overset {\circ}{\Delta}  , n \in \mathbb{Z} \}$, imaginary roots are $ \{ n \delta| n \in \mathbb{Z}, n\neq 0 \}$, with multiplicity $l$. Here $\delta=\sum_{i=0}^{l}a_i \alpha_i $,  
 $\overset {\circ}{\Delta} = \Delta  \bigcap \overset {\circ} {\mathfrak{h}^*}$, $\overset {\circ} {\mathfrak{h}^*}$ is linear span of $\alpha_1,...,\alpha_l$.

As is clear from the above, we need these roots  not for initial $g(A)$, but for dual algebra $g(A^T)$. Then real roots are $ \{ \alpha +n K| \alpha \in \overset {\circ}{\Delta^{\vee}}  , n \in \mathbb{Z} \}$, imaginary roots are $ \{ n K| n \in \mathbb{Z}, n\neq 0 \}$, with multiplicity $l$. Here

\begin{eqnarray}
K=\sum_{i=0}^{l} a_i^{\vee} \alpha_i^{\vee}
\end{eqnarray}
is central element of initial affine Kac-Moody algebra $g(A)$. Positive roots are those with $n > 0$ in above formulae, and, at $n=0$ in the case of real roots, those with $\alpha > 0$.

For calculation of $q$-dimension we need $\langle \lambda_k +\rho,\alpha \rangle = \langle \bar {\rho} +(k+h^{\vee})\Lambda_0,\alpha \rangle  $. We have $\langle \Lambda_0, \alpha \rangle=0$ for  $ \alpha \in \overset {\circ}{\Delta^{\vee}}$, and  $\langle \Lambda_0, K \rangle=1$, since $\alpha_0^{\vee}=1$ for all algebras. Now we can calculate all contributions into (\ref{qd}). 

Contribution of positive real roots with $n=0$ is 1, since numerator and denominator cancel. Contribution of remaining positive real roots  is

\begin{eqnarray}
\Pi_1=\prod_{n=1,2,...}\prod _{\alpha \in \overset {\circ}{\Delta^{\vee}}} \left( \frac{1-q^{\langle\bar{\rho},\alpha \rangle +(k+h^{\vee})n}}{1-q^{\langle\bar{\rho},\alpha \rangle+h^{\vee}n}} \right) 
\end{eqnarray}

Contribution of positive imaginary roots  is
\begin{eqnarray}
\Pi_2=\prod _{n=1,2,...} \left( \frac{1-q^{(k+h^{\vee})n}}{1-q^{h^{\vee}n}} \right)^{l}
\end{eqnarray}

Altogether $q$-dimension is

\begin{eqnarray}
\text{dim}_q(\lambda_k)= \Pi_1 \Pi_2 
\end{eqnarray}

Let's transform this in a way, similar to the well-known transformation of Chern-Simons partition function in the proof of  its duality with topological string. We have $k+h^{\vee}=y, h^{\vee}=t, x=\ln q$:

\begin{eqnarray}
\ln \Pi_1=\sum _{n=1,2,..., \,\alpha \in \overset {\circ}{\Delta^{\vee}}} \left( \ln \left(1-q^{yn+ \langle\bar{\rho},\alpha\rangle}\right) -\ln \left(1-q^{\langle \bar{\rho},\alpha\rangle+tn} \right) \right)= \\ \nonumber
\sum _{p=1,2,...n=1,2,..., \,\alpha \in \overset {\circ}{\Delta^{\vee}}} \frac{1}{p}\left( q^{ptn+ p\langle \bar{\rho},\alpha \rangle}- q^{py n+ p\langle\bar{\rho},\alpha\rangle}\right)\\
\ln \Pi_2=\sum _{n=1,2,...} \left(l \ln \left(1-q^{yn}\right) - l\ln \left(1-q^{tn} \right) \right)= \\ \nonumber
\sum _{p=1,2,...n=1,2,...} \frac{l}{p}\left( q^{ptn}- q^{pyn}\right) 
\end{eqnarray}

Altogether:

\begin{eqnarray}
\ln (\Pi_1\Pi_2)=\sum _{p=1,2,...n=1,2,...} \frac{1}{p}\left( q^{ptn}- q^{pyn}\right) \left(l+  \sum _{\alpha \in \overset {\circ}{\Delta^{\vee}}}  q^{p\langle\bar{\rho},\alpha\rangle} \right)\\ 
\ln (\Pi_1\Pi_2)=\sum _{p=1,2,...n=1,2,...} \frac{1}{p}\left( q^{ptn}- q^{pyn}\right) f(px)=\\  \nonumber
\sum _{p=1,2,...} \frac{1}{p}\left( \frac{1}{1-q^{pt}}- \frac{1}{1-q^{p\delta}}\right) f(px)
\end{eqnarray}
where in the last lines, only, we assume the algebra is of  $\hat{A} \hat{D}\hat{E}$ type.

So, we obtain for $q$-dimension of $k\Lambda_0$ the following expression:

\begin{eqnarray}
\ln \text{dim}_q(\lambda_k)=\sum _{p=1,2,...} \frac{1}{p}\left( \frac{1}{1-q^{pt}}- \frac{1}{1-q^{py}}\right) f(px) =\\  \label{main}
\sum _{p=\pm 1,\pm 2,...} \frac{1}{2p}\left( \frac{1}{1-q^{pt}}- \frac{1}{1-q^{py}}\right) f(px)
\end{eqnarray}
since the summand is even function of $p$.

\section{Partition function of Chern-Simons on 3d sphere as renormalized dimension}

Evidently, (\ref{main}) is the finite sum approximation of the integral: 

\begin{eqnarray} \label{q-int}
 \frac{1}{2} \int^{\infty}_{-\infty} \frac{dx}{x} f(x) \left( \frac{1}{e^{xy}-1} -  \frac{1}{e^{xt}-1} \right)
\end{eqnarray}

This is our main observation: integrand in (\ref{q-int}) (i.e. the summand in (\ref{main}))coincides exactly with that in the integral representation of Chern-Simons free energy (\ref{totalfree2}), taken in the minimal normalization of universal parameters. 

The limit of finite sum to continuous integral corresponds to the limit $q\rightarrow 1$. 

This integral diverges due to singularity at $x=0$. It can be  renormalized by slight deformation of integration contour from real axis  $R$ to $R_+$ or $R_-$. Then we get exactly partition function of Chern-Simons theory on 3d sphere (\ref{totalfree2}).

\section{Conclusion}

In above we found  exact relation between Chern-Simons theory's partition function on 3d sphere and $q$-dimension of certain representation of corresponding affine Kac-Moody algebra. Namely, integrand in universal integral representation of Chern-Simons' free energy (\ref{totalfree2}) coincides  with summand in the infinite sum (\ref{main}), representing logarithm of $q$-dimension. In the limit $q\rightarrow 1$ finite sum becomes integral, although divergent one, which however coincides with (\ref{totalfree2}) after arbitrary small deformation of integration contour around point $x=0$. 

This simultaneous regularization and renormalization by deformation of integration contour is in general agreement with regularization of plethystic sums, suggested in e.g. (\cite{NN}, 5.10), since both lead to answers in terms of multiple Barnes' gamma functions and multiple sine functions, see for Chern-Simons theory \cite{M13,M13b,M14,KM}, for supersymmetric Yang-Mills theory e.g. \cite{NN}, \cite{Oku}. 

It is interesting, that integral representation (\ref{totalfree2}), which proves its relevance in establishing non-perturbative duality of Chern-Simons theory with topological strings \cite{KM}, appears to be relevant in establishing direct connection with  $q$-dimensions theory of affine Kac-Moody algebras, also. 

There are several possible directions of research, in connection with this observation. 

One of them may be calculation of $q$-dimension for other highest-weight representations.  On the Chern-Simons theory side, possible partition functions to be compared are those for Chern-Simons theory on different ("simple") 3d manifolds: 3d torus, spheres product, etc. As a (trivial) example one can consider manifold $S^1 \times S^2$. Its Chern-Simons partition function is 1, which coincides with character and $q$-dimension of a trivial representation of affine Kac-Moody algebra. We also calculate $q$-dimension of the representation with highest weight $m_0\Lambda_0+\sum_{i=1}^{l} \Lambda_i m^i $, where $\Lambda_i, i=0,1,..l$ are fundamental weights of untwisted simply-laced affine Kac-Moody algebra and $m_i, i=0,1,..l$ are integers such that $m_i, i=1,..l$ give a decomposition of highest-weight of adjoint representation of underlying simple Lie algebra w.r.t. its fundamental weights. Level of this representation is $k=m_0+2$, and we again denote $y=k+t$. Then its $q$-dimension is the product of three factors: one is equal to $q$-dimension already calculated (and coinciding with partition function of Chern-Simons theory on 3d sphere), (\ref{main}), second one is  $q$-dimension of corresponding simple Lie algebra, which doesn't depend on coupling $y$, and finally logarithm of the third one is 

\begin{eqnarray} \label{main2}
\sum_{p=1,2,...} \frac{1}{p} \frac{1}{1-q^{-py}} \left(\chi_{ad}(p x(\rho+\theta))-f(p x) \right)
\end{eqnarray}
where $\chi_{ad}(p x(\rho+\theta))$ is character of adjoint representation on the line $ x(\rho+\theta) $, where $\theta$ is the highest-weight of adjoint representation. If instead of $\theta$ we take an arbitrary positive weight $\lambda$, we shall obtain the same formula (\ref{main2}) for $q$-dimension with $\lambda$ instead of $\theta$.  The question is whether one can interpret $q\rightarrow 1$ limit of (\ref{main2}), after appropriate regularization/renormalization, as free energy of Chern-Simons theory on some three-dimensional manifold, normalized by the partition function on 3d sphere. 

Another direction is the following. Taking into account that initial quantity for partition function of Chern-Simons theory on  3d sphere is given by purely Lie-algebraic quantity $S_{00}$ \cite{Wit1}, we obtain some (strange) connection of this and the other purely Lie-algebraic expression - $q$-dimension of representation $k\Lambda_0$. This last quantity, however, has an advantage that it can be generalized to Kac-Moody algebras other than affine ones. One can have in mind "extended" $G^{++}$ and "very extended" $G^{+++}$ Kac-Moody algebras \cite{Nic,W1,W2}, such as $E_{10},E_{11}$, etc. They are suggested as part of description of M-theory, very different from known established approaches.  Dynkin diagrams of extended algebras $G^{++}$ can be obtained \cite{W2} from affine ones, denoted in this notation as $G^{+}$ (i.e. extension of simple Lie algebra $G$), by addition of one node, connected to affine node by simple line. In the same way, Dynkin diagram of very extended algebra $G^{+++}$ appears to be the similar extension of diagram for extended algebra $G^{++}$. So, in each step we have a special node, which probably will play a role of affine node, i.e. one can take corresponding highest-weight fundamental representation and calculate its $q$-dimension by general formula (\ref{qd}), valid for any Kac-Moody algebra. The problem is in that there is no complete description of all roots of these extended Kac-Moody algebras. This doesn't allow us to carry on calculations similar to those in the present paper, however, one can calculate some approximation to $q$-dimension of given extended or very extended algebra, since roots  on the first few levels are completely classified.

\section{Acknowledgments}

I'm grateful to T.Dimofte, S. Garoufalidis and R.Kashaev for discussions of present work. I'm also indebted to organizers of workshop "Low-dimensional topology and number theory" at Oberwolfach, Germany, August 2017, for invitation. Work is partially supported by the Science Committee of the Ministry of Science and Education of the Republic of Armenia under contract  15T-1C233.

\end{document}